\documentclass[11pt]{article}

\usepackage[a4paper,margin=1in]{geometry}
\usepackage[T1]{fontenc}
\usepackage{lmodern}
\usepackage{microtype}
\usepackage{amsmath,amssymb}
\usepackage{graphicx}
\usepackage{booktabs}
\usepackage{array}
\usepackage{tabularx}
\usepackage{xcolor}
\usepackage{tikz}
\usetikzlibrary{arrows.meta,positioning,shapes.geometric,backgrounds,fit}
\usepackage{caption}
\usepackage[hidelinks]{hyperref}
\hypersetup{colorlinks=true,linkcolor=black,citecolor=black,urlcolor=blue!55!black}

\definecolor{bandgray}{gray}{0.90}
\definecolor{da}{RGB}{201,127,23}
\definecolor{pigreen}{RGB}{31,128,82}
\definecolor{dshblue}{RGB}{44,95,158}
\definecolor{gapline}{RGB}{170,20,20}

\newcommand{\da}{\textsc{deepagents}}
\newcommand{\pipi}{\textsc{pi}}
\newcommand{\dsh}{\textsc{dsh}}
\newcommand{\code}[1]{\texttt{\small #1}}

\definecolor{convgreen}{RGB}{34,110,60}
\definecolor{convamber}{RGB}{196,126,20}
\newcommand{\yes}{{\color{convgreen}\checkmark}}
\newcommand{\nay}{{\color{gapline}$\times$}}
\newcommand{\prt}{{\color{convamber}$\blacktriangle$}}

\title{\vspace{-2.5em}\bfseries The Empire, Long Divided, Must Unite:\\ Architectural Convergence in Three LLM Agent Harnesses\thanks{The title adapts the opening line of Luo Guanzhong's \emph{Romance of the Three Kingdoms} (trans.\ M.~Roberts): ``The empire, long divided, must unite; long united, must divide.'' Here three harnesses, from opposing philosophies, unite on one middle form, and on one dimension (Section~\ref{sec:gap}) they remain, for now, divided.}}
\author{Jiahong Dai\\[2pt]\normalsize \texttt{jiahong001@e.ntu.edu.sg}}
\date{25 August 2026}

\begin{document}
\maketitle

\begin{abstract}
\noindent
An \emph{agent harness} is what turns a language model into an autonomous agent: the surrounding code that builds the model's context, mediates its tools, runs the loop, and persists state across a long-horizon run. This layer, not the model it wraps, is increasingly the binding constraint on agent behaviour. We present a source-level, multi-case study of three open coding-agent harnesses built from deliberately opposing philosophies: LangChain's \da{} (batteries-included), Earendil's \pipi{} (radical minimalism), and DeepSeek's \dsh{} (everything-is-a-plugin). Reading each at a pinned commit and following its commit history, we find that the two mature harnesses have travelled in \emph{opposite} directions (\da{} subtracting authored scaffolding, \pipi{} accreting durable infrastructure), yet converged toward one architectural middle form of five recurring elements: a commoditised loop, an append-only replayable session record, model quirks kept as data, progressive disclosure of context, and explicit extension seams. A third harness, read afterward as a held-out check, exhibits all five, and in one seam reuses another's implementation outright. We therefore do not claim independent invention, and decompose the convergence into parallel discovery, diffusion, and literal reuse. Finally, one load-bearing dimension shows no convergence, and indeed no presence: \emph{external verifiability}, a tamper-evident record an outside party can check without trusting the runtime. We read this absence not as an oversight but as a predictive gap, the next axis on which harnesses for provenance-sensitive domains will differ.
\end{abstract}


\section{Introduction}
Autonomous coding agents have advanced quickly on long-horizon software tasks~\cite{sweagent,codeact,autogen}, and the progress is usually credited to the model. A growing body of evidence points elsewhere. Among models of comparable frontier capability, the \emph{harness} that surrounds the model, not the model itself, often governs the larger share of performance variance on long-horizon tasks. Recent work formalises this as a ``binding constraint'': single-harness changes move Terminal-Bench~2 pass@1 by several points and SWE-bench Verified by up to fifteen, with the model held fixed~\cite{disclose}. If the harness is the binding constraint, then where harness architectures are heading is a first-order question, not an implementation detail.

Yet prior work maps this layer only partially. The nearest source-level taxonomy~\cite{scaffold} reads thirteen coding-agent scaffolds at pinned commits, but excludes \da{} and analyses static snapshots only. A longitudinal study of five command-line harnesses~\cite{dontblame} relates release velocity to quality, but not to the architectural form that evolution produces. And the survey and reading list that treat the harness as a research object~\cite{survey1,survey2} do not read implementations at all.

Three gaps therefore remain in our understanding of the harness layer. First, \textbf{the harness layer itself is unread} (G1), since existing source-level studies cover coding-agent applications rather than the general harness layer beneath them, and neither \pipi{} nor \dsh{} appears in any architectural study. Second, \textbf{architectural trajectories are uncharacterised} (G2), since prior readings analyse static snapshots and the one longitudinal study quantifies quality rather than form. Third, \textbf{convergence is neither claimed nor tested} (G3), since the only study to observe converging dimensions reports the observation as incidental and asks nothing about where the field's architectures are heading.

To address these gaps, we present a source-level, multiple-case study of three open coding-agent harnesses chosen for maximal philosophical spread:
\begin{itemize}\itemsep2pt
\item \da{} (LangChain): batteries-included, with a middleware stack, pluggable storage backends, first-class subagents, and four layers of automatic context management, wired before the developer writes a line.
\item \pipi{} (Earendil Works; Mario Zechner and Armin Ronacher): the ``smallest useful harness,'' with four tools, a $\sim$300-token base prompt, and the credo that everything else is an extension the user can see.
\item \dsh{} (DeepSeek): everything-is-a-plugin, a service-locator runtime in which even the agent loop is a swappable configuration row.
\end{itemize}

\begin{figure}[t]
\centering
\includegraphics[width=0.7\linewidth]{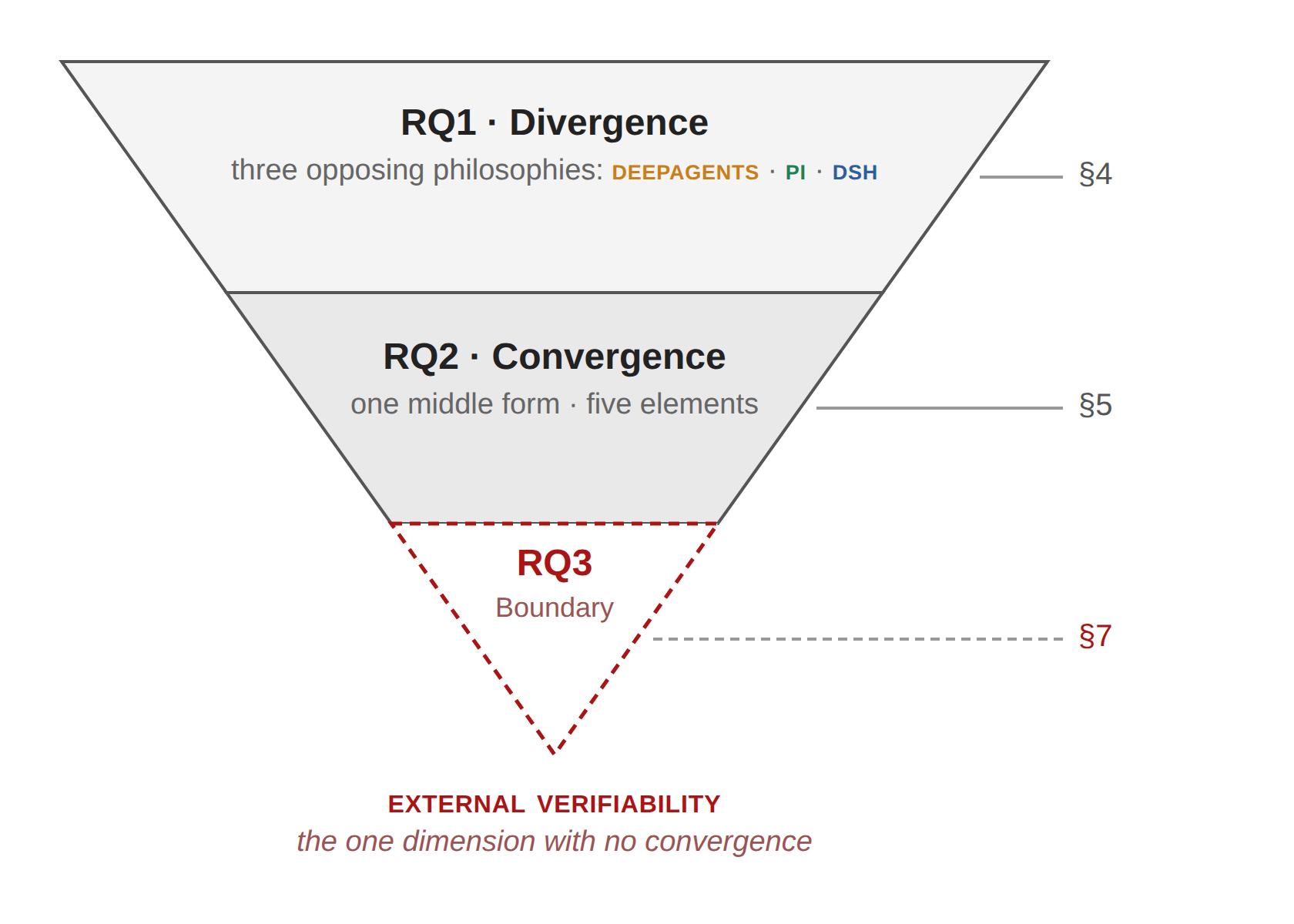}
\caption{\textbf{The three research questions.} Each band narrows to the next: three origins (RQ1, Section~\ref{sec:origins}), one convergent form (RQ2, Section~\ref{sec:form}), one open dimension (RQ3, Section~\ref{sec:gap}). Dashed\,=\,open, as in every figure.}
\label{fig:rqs}
\end{figure}

Reading all three does not end in taking a side. Close reading, close enough to reproduce defects and file issues in two of the trackers, surfaces a different picture. We organise the study around three research questions (RQs), each answered by a tagged section (Sections~\ref{sec:origins}, \ref{sec:form}, and \ref{sec:gap}):
\begin{description}\itemsep2pt
\item[RQ1 (Divergence).] From what architectural positions did the harnesses start, and along what trajectories have they evolved?
\item[RQ2 (Convergence).] Is there a common form they are arriving at; what is it; and \emph{why} are they converging?
\item[RQ3 (Boundary).] Is there a dimension on which they have \emph{not} converged, and what explains the gap?
\end{description}
Figure~\ref{fig:rqs} charts how the questions narrow: from three origins, to one convergent form, to the one dimension that remains open.

Our central finding is the following:
\begin{quote}\itshape
Consistent with a shared and intensifying selection pressure (long-horizon autonomous operation), three harnesses that began from opposing philosophies have converged on one architectural middle form; the convergence is real but not fully independent, and its boundary, the one dimension where no convergence has occurred, is external verifiability.
\end{quote}

\paragraph{Contributions.} Our main contributions are fivefold:
\begin{itemize}\itemsep2pt
\item \textbf{Source-Level Coverage} (G1): the first source-level architectural study of the harness layer represented by \da{}, \pipi{}, and \dsh{}; the nearest prior taxonomy explicitly excludes \da{}, and \pipi{} and \dsh{} appear in no prior architectural study.
\item \textbf{Trajectory Evidence} (G2): the first account of harness \emph{trajectories} from commit archaeology (subtraction, accretion, and reuse), where prior work is either a static snapshot or a quality-focused longitudinal study.
\item \textbf{The Convergent Middle Form} (G3): a five-element middle form grounded in two cases, checked against a third read subsequently (not fully independent of the second), and decomposed into parallel discovery, diffusion, and literal reuse.
\item \textbf{Convergent Fault Lines}: an illustrative taxonomy of four architectural seams on which reproduced defects recur across opposing designs.
\item \textbf{The Verifiability Boundary}: evidence that \emph{external verifiability} is a dimension on which all three harnesses are absent, and the argument that its absence is a predictive gap rather than an oversight.
\end{itemize}
To the best of our knowledge, this is the first study to make architectural convergence the central, evidenced claim about the LLM agent harness layer: five recurring elements, three convergence mechanisms, and one dimension on which convergence has not begun.

Three scope caveats bound these claims. We do not claim the three converged independently: they are mutually aware, and one literally reuses another (Section~\ref{sec:mech}). We do not claim a systematic longitudinal analysis; commit trajectories are directional corroboration, not a controlled study. And we do not use this paper to advance any particular solution to the verifiability gap: we characterise the gap, and stop.

\begin{table}[!ht]
\centering
\caption{\textbf{Capability matrix: three production harnesses vs.\ the prescriptive protocol layer.} \yes: present, \prt: partial, \nay: absent, ---: not applicable. \emph{Protocols} scores what prescriptive proposals (Autogenesis~\cite{autogenesis}, MCP, A2A) \emph{specify}, not what any shipped harness implements.}
\label{tab:matrix}
\small
\renewcommand{\arraystretch}{1.22}
\begin{tabularx}{\linewidth}{@{}>{\raggedright\arraybackslash}X ccc @{\hspace{2em}} c@{}}
\toprule
 & \multicolumn{3}{c}{\textbf{Built bottom-up}} & \textbf{Proposed top-down} \\
\cmidrule(lr){2-4} \cmidrule(l){5-5}
\textbf{Capability} & \textbf{\da{}} & \textbf{\pipi{}} & \textbf{\dsh{}} & \textbf{Protocols} \\
\midrule
\multicolumn{5}{@{}l}{\itshape Convergent middle form (\S\ref{sec:form})} \\
Commoditised loop (small, readable, or delegated) & \yes & \yes & \yes & --- \\
Append-only, replayable session record        & \prt & \yes & \yes & \prt \\
Model quirks externalized as data             & \yes & \yes & \yes & \prt \\
Progressive disclosure of context             & \yes & \yes & \yes & \prt \\
Orthogonal extension seams                    & \yes & \yes & \yes & \yes \\
\midrule
\multicolumn{5}{@{}l}{\itshape Residual divergence (\S\ref{sec:form})} \\
OS-enforced sandbox, per-call policy          & \prt & \nay & \yes & \nay \\
\midrule
\multicolumn{5}{@{}l}{\itshape The verifiability gap (\S\ref{sec:gap})} \\
Externally verifiable, tamper-evident record  & \nay & \nay & \nay & \prt \\
Domain provenance (as-of, entitlement, cost)  & \nay & \nay & \nay & \prt \\
\bottomrule
\end{tabularx}
\end{table}

As Table~\ref{tab:matrix} shows, the comparison separates a convergent core from a divergent gap. The five middle-form capabilities are present across all three production harnesses (with the append-only record only partially realised in \da{}, Section~\ref{sec:form}), yet the prescriptive protocols specify them only partially, the seam row excepted, which protocols exist to standardise. The middle band holds the one capability that remains genuinely divergent: an OS-enforced sandbox ships in \dsh{}, partially in \da{}, and deliberately not at all in \pipi{} (``containerise yourself''), the audience split of Section~\ref{sec:form}. The verifiability rows invert this: absent from every production harness, but placed at the centre of prescriptive proposals such as Autogenesis, which specifies runtime-internal lineage, itself short of outsider-checkable, hence a partial mark. That asymmetry is the paper's boundary claim: what practitioners converged on bottom-up, protocols only partly specify; and the verifiable record protocols prescribe top-down, no shipped harness has built.

\section{Background and Related Work}\label{sec:related}
\paragraph{Harness engineering as an object of study.} The harness has recently been recognised as a distinct research object, with a survey formalising it as a tuple of execution loop, tool registry, context manager, state store, lifecycle hooks, and evaluation interface~\cite{survey1}, a curated reading list mapping the area~\cite{survey2}, and a subfield emerging around automatic harness optimisation~\cite{disclose}. These works establish the vocabulary but do not read implementations comparatively; a comparative, source-level reading of the harness layer is precisely what this paper supplies.

\paragraph{The nearest neighbour.} Closest to our work is a source-code taxonomy of thirteen coding-agent scaffolds at pinned commits, characterising each across twelve dimensions in three layers, grounded in file and line references~\cite{scaffold}. We inherit its methodological discipline (pinned commits, line-level evidence) and differ in three ways that define our contribution. First, \emph{objects}: that study explicitly excludes \da{} as ``general-purpose rather than coding-specific,'' and neither \pipi{} nor \dsh{} appears in it; we study the general harness layer beneath coding-agent applications. Second, \emph{time}: it analyses static snapshots and disavows evolution tracking; our central evidence is the trajectory each harness has travelled. Third, \emph{framing}: it reports convergence as an incidental observation (``dimensions converge where external constraints dominate'') and treats persistence as a detail within state management; we make convergence the central thesis and treat auditability as a first-class dimension. A complementary longitudinal study tracks five command-line harnesses over twelve months and finds that rapid release velocity does not yield proportional quality gains~\cite{dontblame}; it quantifies \emph{evolution-to-quality} but does not characterise the \emph{architectural form} that evolution is producing, which is our subject. In short, prior work draws a static map of the design space; this paper traces three trajectories through it and identifies the point at which they meet.

\paragraph{Why the destination matters.} That the harness can dominate the model motivates studying its architecture directly~\cite{disclose}. The insight that the agent-computer interface shapes capability originates with tool-centric agents~\cite{sweagent,codeact}; how tools are organised and exposed further shapes agent behaviour~\cite{devil}, and scaffolding can determine performance more than model choice~\cite{confucius}; and multi-agent frameworks established the harness as reusable infrastructure~\cite{autogen}. This line establishes that the interface shapes capability; it does not ask where harness architectures are heading, which is the question this paper answers.

\paragraph{The prescriptive counterpart.} A complementary line approaches the harness top-down, by specifying what it \emph{should} provide: MCP standardises tool invocation and A2A agent-to-agent messaging~\cite{mcp,a2a}, and Autogenesis~\cite{autogenesis} (part of a broader self-evolving-agents programme~\cite{selfevolve}) proposes a protocol above both that registers prompts, agents, tools, environments, and memory as versioned, lifecycle-managed resources, so that every self-modification is traceable and reversible. We work in the opposite direction: rather than prescribing what harnesses should look like, we document at the source level what three production harnesses have in fact become, and the auditable lineage such proposals place at their centre is precisely the dimension on which all three show no convergence, and no presence (Section~\ref{sec:gap}).

\paragraph{Method lineage.} We follow established guidance for multiple-case study design and reporting in software engineering~\cite{yin,runeson}, and treat source-level architectural narrative in the tradition of open-source architecture studies~\cite{aosa}. These sources supply the method; none applies it to the agent harness, the object this study reads.

\section{Study Design}\label{sec:design}
\paragraph{Design.} This is an explanatory, theory-building multiple-case study following a \emph{literal replication} logic across maximum-variation cases: the first two cases (\da{}, \pipi{}) ground a candidate model of the convergent form, and the third (\dsh{}) is read subsequently as a held-out check that, because \dsh{} reuses \pipi{}'s provider catalogue, is confirmatory only for the elements where its instantiation is independent (Section~\ref{sec:mech}). Sections~\ref{sec:origins}, \ref{sec:form}, and \ref{sec:gap} answer RQ1, RQ2, and RQ3 respectively; Section~\ref{sec:seams} adds an illustrative second evidence line for RQ2.

\paragraph{Case selection.} Cases were chosen for maximal spread, not convenience. They span three points of the design space (batteries-included / minimal / everything-is-a-plugin), three organisational forms (a venture-backed framework company / two independent engineers / a frontier lab), and two languages (Python / TypeScript). Each is open-source and readable at a pinned revision. The candidate pool was open-source, actively developed, source-readable coding-agent harnesses; from it we selected for maximal philosophical spread. Prominent harnesses we did not read at source level (OpenHands, Aider, SWE-agent's harness, Gemini CLI, Codex CLI, Claude Code) are not counted as evidence; a documentation-level scan of two of them for the five elements is future work (Section~\ref{sec:threats}), and where Table~\ref{tab:matrix} and the text say ``the field,'' the claim is scoped to these three cases. We deliberately \emph{disclose an independence limitation}: \dsh{}'s generic provider adapter depends on \pipi{}'s published package (Section~\ref{sec:mech}); the three are therefore not fully independent lineages, which we handle analytically rather than assume away.

\paragraph{Pinned revisions.} All file and line references are to \da{}~0.7.8 at commit \code{2c8015378} (trajectory evidence also cites \code{23b83ad50}), \pipi{} \code{main}\,@\,\code{a470b121b}, and \dsh{} at release tag \code{dsh-v0.1.1-rc.2}, commit \code{b150a551b}.\footnote{Repositories: \url{https://github.com/langchain-ai/deepagents}, \url{https://github.com/earendil-works/pi}, \url{https://github.com/deepseek-ai/deepseek-harness}.} Line counts are whole-file, blank and comment lines included.

\paragraph{Data sources.} Four kinds, all re-checkable: (i)~\emph{source} at the pinned revisions, with claims anchored to file:line; (ii)~\emph{commit, PR, and issue archaeology} for trajectories; (iii)~\emph{hands-on reproduction}, comprising two plugins we wrote against \dsh{}'s runtime, and two defects (the \da{} path-normalisation bypass and the \pipi{} throttle-misclassification) independently re-derived in a sandbox, with further seam instances located by source reading; and (iv)~\emph{upstream confirmation of specific claims}. On (iv): we submitted issues and a pull request; one documentation fix was accepted, assigned, and merged, and one defect was filed (issue~\#5640). These confirm individual point claims (not the five-element model or the seam taxonomy), and we scope their evidential weight accordingly (Section~\ref{sec:threats}).

\paragraph{A priori dimensions.} The comparison dimensions (Table~\ref{tab:dims}) were fixed before close reading, to avoid fitting a framework to observations. Where they overlap the nearest neighbour's twelve dimensions we adopt its discipline; we add two axes it lacks (\emph{auditability} and \emph{evolution}) and refine two (session-record \emph{strength}, recovery \emph{semantics}) from binary presence to graded form.

\paragraph{Reading with AI assistance.} The source sweep used parallel AI agents to locate and excerpt code; every claim was then verified by hand against the source. We disclose this because it bears on reliability, and because the reproduction artefacts (below) are the durable check on it. Reading depth is asymmetric and we report it as such: \da{} and \pipi{} were read line-by-line; \dsh{} received one thorough pass plus a hands-on plugin experiment.

\section{Divergent Origins (RQ1)}\label{sec:origins}
\begin{figure}[t]
\centering
\includegraphics[width=\linewidth]{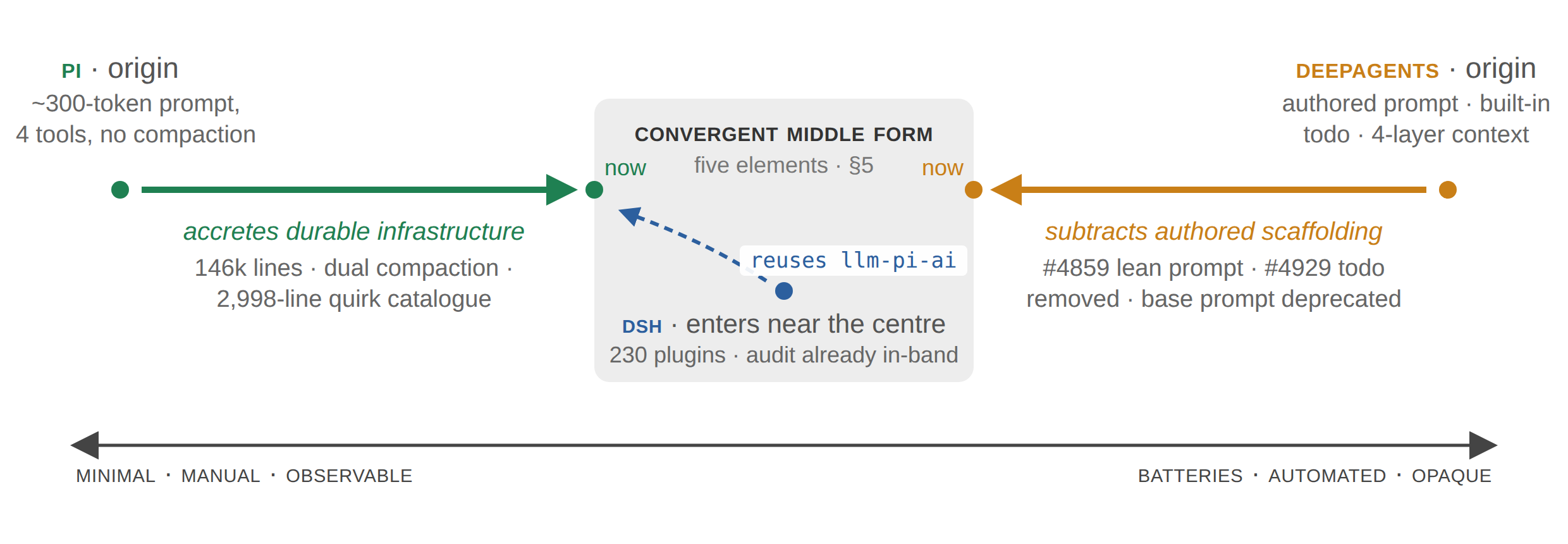}
\caption{\textbf{Divergent origins, converging trajectories.} Positions are interpretive, not measured; arrows denote the direction of each project's commit history, not speed. \da{} subtracts authored scaffolding; \pipi{} accretes durable infrastructure; \dsh{} enters near the middle and, in one seam, literally reuses \pipi{}.}
\label{fig:diverge}
\end{figure}

This section answers RQ1. As shown in Figure~\ref{fig:diverge}, the three harnesses occupy a single spectrum (how much the harness automates versus how much it leaves observable) and travel it in three ways: i)~\da{} subtracts authored scaffolding from the batteries-included end; ii)~\pipi{} accretes durable infrastructure from the minimal end; and iii)~\dsh{} enters near the centre and, in one seam, literally reuses \pipi{}.

\subsection{\da{}: the subtracting maximalist}
\textbf{Position.} \da{} is an assembly layer over LangChain's runtime: planning, filesystem, subagents, summarisation, memory, skills, permissions, and human-in-the-loop are \emph{all} middleware, composed in a three-phase stack (\code{graph.py:816--893}).

\textbf{Architecture.} Storage is a backend protocol whose methods default to \code{NotImplementedError}, so capabilities are optional and probed by identity comparison. Context is defended in four automatic layers: a delta-channel reducer (replacing LangGraph's quadratic whole-list checkpointing), LLM summarisation near 85\% of the window, eviction of oversized tool results to storage behind a head/tail preview and a pointer, and cheap truncation of stale tool arguments first. Subagent isolation is careful: the \code{task} tool copies parent state minus messages and private fields inward and returns only a final message outward, with private fields late-bound after assembly.

\textbf{Trajectory.} \da{} subtracts. The commit history records a trajectory the project's public positioning does not: the authored base prompt is deprecated with a note that the harness ``no longer provides an authored base prompt.'' PR~\#4859 stripped built-in tool-usage prose (``lean system prompt by default''). PR~\#4929 removed the todo/planning middleware from the default stack, leaving it only as an opt-in within one model's profile: planning reclassified from a harness default to a per-model need. Each step moves toward \pipi{}'s stated positions.

\subsection{\pipi{}: the accreting minimalist}
\textbf{Position.} \pipi{}'s public identity is minimalism~\cite{pipost}, and the loop honours it: its agent package is 2{,}368 lines (the loop file, \code{agent-loop.ts}, is 796), readable end-to-end, with no \code{try}/\code{catch} in the loop at all (Section~\ref{sec:form}).

\textbf{Architecture.} Beneath the minimal loop, \pipi{} has built serious infrastructure: a 2{,}941-line formal specification defining three durable forms (immutable entries, mutable registers, append-only usage rows) plus a durable ``program counter'' for crash recovery that overwrites a single register per operation, so recovery reads the register and resumes without replaying the log.

\textbf{Trajectory.} \pipi{} accretes. The repository around the loop is 146{,}170 lines across ten packages. The harness that rejected subagents now reserves ``lanes'' in its session model naming subagents as a design target; the one that rejected compaction runs two implementations mid-migration; the one that dismissed per-model coaching maintains a 2{,}998-line generator producing one of the most detailed model-quirk catalogues in the open, with roughly fifteen compatibility flags for a single wire format. What it rejected in prose returned as data.

\subsection{\dsh{}: the plugin absolutist}
\textbf{Position.} \dsh{} is a TypeScript monorepo of $\sim$230 plugins over a vendored service-locator runtime.

\textbf{Architecture.} A plugin is any object implementing a service; consumers resolve a service by a stable context key and never import an implementation; load order is expressed declaratively via \code{inject}; registration is a reversible side effect that unwinds on reload. Even the agent loop is a configuration row (\code{ctx.agentLoop}), replaceable from config: ``everything is a plugin'' holds literally. Its session model is the strongest audit story of the three (Section~\ref{sec:form}), and its sandbox throws \code{SANDBOX\_UNAVAILABLE} rather than run unconfined, with policy travelling per call rather than per provider.

\textbf{Trajectory.} \dsh{} reuses. It did not travel far to reach the middle; it entered near it. Its generic provider adapter, \code{llm-pi-ai}, depends directly on \pipi{}'s published \code{@earendil-works/pi-ai} package and is described in-tree as a ``design-verification twin,'' a case we return to in Section~\ref{sec:mech}.

\section{The Convergent Form (RQ2)}\label{sec:form}
\begin{figure}[t]
\centering
\includegraphics[width=\linewidth]{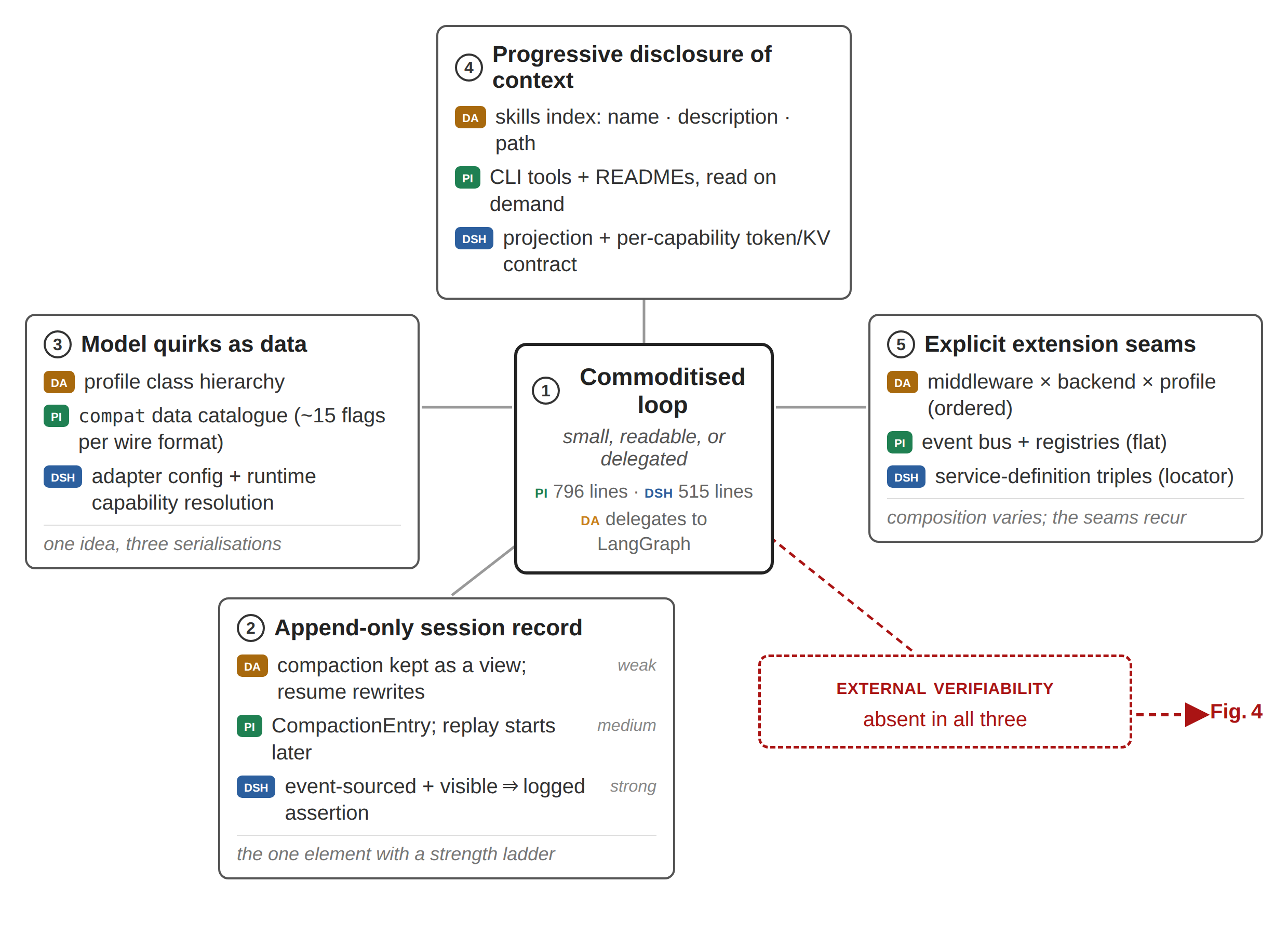}
\caption{\textbf{The convergent middle form.} One idea, three serialisations: each element lists its \da{} (da), \pipi{} (pi), and \dsh{} instantiation. Solid boxes are converged elements; the dashed slot is the non-converged dimension of Section~\ref{sec:gap} (Figure~\ref{fig:ladder}).}
\label{fig:form}
\end{figure}

This section answers the first half of RQ2: the destination. As Figure~\ref{fig:form} shows, the form is a commoditised loop with the remaining four elements at its seams, each instantiated three different ways; the dashed slot marks the one dimension where no harness appears (Section~\ref{sec:gap}). Table~\ref{tab:dims} gives the per-dimension evidence for the five elements.

\begin{table*}[t]
\centering
\caption{\textbf{Nine dimensions across three harnesses.} The five convergent elements of Section~\ref{sec:form} are typeset in \textbf{bold}; the final row is the non-converged dimension of Section~\ref{sec:gap}. Abbrev.: da\,=\,\da{}.}
\label{tab:dims}
\small
\renewcommand{\arraystretch}{1.25}
\begin{tabularx}{\textwidth}{@{}l >{\raggedright\arraybackslash}X >{\raggedright\arraybackslash}X >{\raggedright\arraybackslash}X@{}}
\toprule
\textbf{Dimension} & \textbf{\da{}} & \textbf{\pipi{}} & \textbf{\dsh{}} \\
\midrule
Philosophy & batteries-included middleware & minimal loop, extend outward & everything-is-a-plugin \\
\textbf{Loop (commoditised)} & none owned (delegates to LangGraph) & 796-line loop file, no try/catch & 515-line driver, \emph{itself} a swappable config row \\
\textbf{Append-only session} & delta-channel; compaction kept as a \emph{view} (weak) & append-only entries + registers (medium) & event-sourced + runtime ``visible$\Rightarrow$logged'' assertion + replace-op (strong) \\
Crash recovery & checkpoint; resume \emph{rewrites} history & durable register; read-and-resume & \emph{closes} an interrupted turn; refuses mid-stream corruption \\
\textbf{Model quirks as data} & profile class hierarchy & \code{compat} data catalogue (defaults inferred) & adapter config + runtime capability resolution \\
\textbf{Progressive disclosure} & skills index (name/desc/path) & CLI tools + READMEs, read on demand & projection + per-capability token/KV contract \\
\textbf{Explicit seams} & middleware / backend / profile (ordered) & event bus + registry (enumerated) & service-definition triples (locator) \\
Sandbox & permissions + HITL + paid remote backends & none (``YOLO''; containerise yourself) & bwrap/Landlock; throws rather than degrade \\
\midrule
\textbf{\textcolor{gapline}{External verifiability}} & \textcolor{gapline}{none (outsourced telemetry)} & \textcolor{gapline}{none (mutable private files)} & \textcolor{gapline}{operational only (not third-party checkable)} \\
\bottomrule
\end{tabularx}
\end{table*}

\subsection{Five recurring elements}
\textbf{(1) A commoditised loop.} \pipi{}'s agent-loop file is 796 lines; \dsh{}'s driver is 515; \da{} does not own a loop at all, delegating to LangGraph. The element is not loop size but that none of the three competes on the loop: it is small, readable, or delegated wholesale, never a locus of differentiation, a claim a harness built around a large proprietary loop would falsify. A \pipi{} detail worth adopting is the never-throw stream contract; provider failures become a terminal assistant message with \code{stopReason} of \code{error} or \code{aborted}, and on \code{stopReason == length} \emph{every} tool call in the message is failed, since any may carry truncated JSON.

\textbf{(2) An append-only, replayable session record}, and here the three form a strength ordering. \da{} keeps compaction as a \emph{view} over an untouched history (``for replay, evals, and shared state,'' says the source), which is weak because recovery still rewrites messages. \pipi{} appends a \code{CompactionEntry} carrying \code{firstKeptEntryId}; nothing is destroyed and replay simply starts later, a medium form (this is \pipi{}'s newer harness path; an older interactive compaction still overwrites history, the defect of Section~\ref{sec:seams}). \dsh{} is strongest: an event-sourced log with a runtime-asserted invariant that anything a model saw must be reconstructable from the log, history a projection rather than storage, and a \code{replace} surface-op that masks a span while preserving the original events. All three learned that mutating history is a trap; they differ in how far they enforce it.

\textbf{(3) Model quirks as data.} \pipi{}'s \code{compat} catalogue, \da{}'s registered profiles, and \dsh{}'s adapter config with runtime capability resolution are one idea in three serialisations: a class hierarchy, a data directory, and a runtime resolver. \dsh{}'s is the most dynamic and, correspondingly, the least inspectable at rest.

\textbf{(4) Progressive disclosure of context.} All three put only a name, description, and path in the prompt and read the body on demand (\da{}'s skills index, \pipi{}'s CLI-tools-with-READMEs, \dsh{}'s runtime projection), and both \da{} and \pipi{} adopted the \code{AGENTS.md} convention and the filesystem as memory. \dsh{} goes furthest, attaching a written per-capability contract stating each plugin's token and KV-cache cost.

\textbf{(5) Explicit seams instead of a monolith.} \da{} exposes three orthogonal axes (middleware, backend, profile); \pipi{} a flatter event bus plus registries; \dsh{} named service-definition triples where one provider swap moves an entire capability world. The disagreement is composition style (ordered middleware versus event bus versus locator), not whether seams should exist.

Three findings emerge from Table~\ref{tab:dims}. First, all five elements recur in all three columns despite opposing philosophies, consistent with the middle form being a property of the problem rather than of any single lineage. Second, no element recurs as the same code (a class hierarchy, a data directory, and a runtime resolver serialise one idea three ways), suggesting convergence of idea rather than of implementation. Third, only the session record forms a strength ladder (view $<$ append-only $<$ event-sourced with a runtime assertion) rather than a cluster, and that ladder points directly at the non-converged dimension of Section~\ref{sec:gap}. The practical implication is direct: treat the five shapes as settled, adopt the strongest available form of each, and spend novelty on the dimensions that remain unsettled.

\subsection{The residual disagreement is audience, not architecture}
Where the three still differ is \emph{how much to automate versus leave observable}. \da{} automates context management because its users embed agents in products; \pipi{} keeps it manual because its user is a person at a terminal who wants to see everything. That is a difference in audience, not in architecture, and it predicts the shape of each project's remaining roadmap without contradicting the convergence.

\subsection{Mechanisms of convergence}\label{sec:mech}
This subsection answers the second half of RQ2: the mechanism. Two teams arriving at the same design is a weak signal: they could have read each other. We therefore decompose the convergence into three mechanisms rather than assert independent invention.

\emph{Parallel discovery} is visible where the same shape is reached from demonstrably different starting prose and different pressures: the append-only record reached by \pipi{} from crash pressure and by \da{} from checkpoint-cost pressure; progressive disclosure reached by both before either shipped subagents. \emph{Diffusion} we concede openly: \pipi{} and \da{} are both public and mutually known, and shared conventions like \code{AGENTS.md} spread by imitation, not rediscovery. \emph{Literal reuse} is the strongest and most concrete: \dsh{} mounts \pipi{}'s provider catalogue wholesale through \code{llm-pi-ai}, one team adopting another's solved problem rather than defending its own. This is convergence in its most concrete and verifiable form, and it is exactly the ``adopt it, don't defend your own version'' move the convergent form recommends, occurring between the projects themselves. The thesis does not require independence; it requires that the same shapes recur under shared pressure, which all three mechanisms support and none undercuts.

\section{Convergent Fault Lines}\label{sec:seams}
Architectures are not the only thing shaped by a shared problem; \emph{defects} may be too. The defects we reproduced or identified across \da{} and \pipi{} (two re-derived in a sandbox, the rest located by source reading) each fall on one of four recurring seams (Table~\ref{tab:seams}). We present these as illustrative case reports, not as an independent test: the four seams were induced from a small, analyst-directed sample, some are language-specific (S1's sync/async drift cannot arise in \pipi{}'s single-threaded model), and S4 is a cross-cutting property rather than a fourth bin. They \emph{suggest}, rather than establish, that these fault lines track the problem shape.

\begin{table}[t]
\centering
\caption{The four seams on which every reproduced defect landed. Loss column: $\bullet$\,=\,irreversible, $\circ$\,=\,recoverable, ---\,=\,n/a. da\,=\,\da{}.}
\label{tab:seams}
\footnotesize
\renewcommand{\arraystretch}{1.25}
\begin{tabularx}{\linewidth}{@{}>{\raggedright\arraybackslash}p{1.85cm} >{\raggedright\arraybackslash}p{2.95cm} >{\raggedright\arraybackslash}X c@{}}
\toprule
\textbf{Seam} & \textbf{Specimen} & \textbf{Mechanism $\rightarrow$ consequence} & \textbf{Loss} \\
\midrule
S1 sync/async drift & da, \code{async\_subagents} & guard inside \code{try} in the sync twin, outside it in the async twin $\rightarrow$ unhandled \code{KeyError} on recovery & $\circ$ \\
\addlinespace
S2 normalisation trust gap & da, \code{utils.py:691} & \code{//secrets} $\neq$ \code{/secrets} by path component but $=$ by filesystem $\rightarrow$ delete-deny, approval, and routing checks fail open & $\bullet$ \\
 & \pipi{}, \code{overflow.ts:75} & throttle rethrown as a bare object, JSON-stringified past a caret-anchored regex $\rightarrow$ destructive compaction on a transient throttle & $\bullet$ \\
\addlinespace
S3 string-matched semantics & \pipi{}, \code{overflow.ts:60} & error meaning decided by ${\sim}25$ regexes $\rightarrow$ a reworded rate-limit misread as context overflow & $\bullet$ \\
\addlinespace
S4 silent vs.\ loud failure & all four specimens & an unrecognised shape defaults and runs on instead of stopping (cf.\ da \code{graph.py}, which raises) & --- \\
\bottomrule
\end{tabularx}
\end{table}

\textbf{S1: Sync/async drift.} In \da{}, a hand-written synchronous tool guards a client lookup inside a \code{try}; its asynchronous twin performs the same lookup outside it (\code{async\_subagents.py:445} and \code{:625}, against the guarded sync halves at \code{:421}/\code{:594}). A checkpointed task that references a since-renamed subagent therefore crashes the async caller with an unhandled \code{KeyError} while the sync caller degrades to a clean error; a third instance survives in the update tool's two branches (\code{:496}/\code{:535}). Paired async written by hand is two copies of one intent that the compiler never checks for agreement. \emph{Rule.} Derive one twin from the other or share a single guarded core; where both are hand-written, diff their guard clauses and test both halves against the same failure. (Filed as issue \#5640.)

\textbf{S2: Normalisation trust gaps.} A value crosses a boundary and the two sides disagree on its shape, with no enforced canonicalisation between. In \da{}, \code{os.path.normpath} preserves exactly two leading slashes (POSIX-reserved), so \code{//secrets} compares unequal to \code{/secrets} by path component (\code{validate\_path}, \code{utils.py:691}; \code{\_paths\_overlap}, \code{:607--616}) while the filesystem strips the redundant slash and resolves both to one file. Three component-based checks fail open: a delete deny-rule (\code{filesystem.py:558}), a human-approval interrupt (\code{\_fs\_interrupt.py:126}), and composite-backend routing (\code{composite.py:166--177}), while the adjacent \code{wcmatch}-based read/write/edit checks (\code{filesystem.py:420--430}), which normalise, are immune. In \pipi{}, a mid-stream Bedrock throttling error is rethrown as a bare object (\code{bedrock-converse-stream.ts:306}), fails an \code{instanceof} guard (\code{:385--390}), is JSON-stringified (\code{error-body.ts:38}), and slips past a caret-anchored negative regex written to catch exactly it (\code{overflow.ts:75})---because the string now begins with a brace. Same disease, two languages. \emph{Rule.} One mandatory canonicalisation point per trust boundary; downstream accepts only the canonical form.

\textbf{S3: Semantics by string-matching.} \pipi{} classifies provider errors with ${\sim}25$ regexes plus a negative list (\code{overflow.ts:60--142}); the Bedrock case is its failure mode. Deciding meaning from text means a reworded or re-wrapped message flips the decision, and here the cost of being wrong is a transient rate-limit routed into destructive compaction, a paid summarisation that permanently replaces conversation history via \pipi{}'s older interactive compaction path (\code{agent-session.ts:2079}), not the append-only \code{CompactionEntry} of Section~\ref{sec:form}; the same path also fails to mark the error retryable (\code{retry.ts:26--44}), so the throttle is neither retried nor survived. \emph{Rule.} Errors carry a typed kind; regex is a last-resort fallback that must fail toward the safe side: here, ``do not compact'' when unsure, since compacting a non-overflow is unrecoverable while declining to compact a real overflow is not.

\textbf{S4: Silent degradation versus loud failure.} Every defect above keeps running after the mistake: the bare object becomes a string and flows on, the \code{//} path routes to the wrong backend, the misclassified throttle proceeds to compaction. Contrast \da{}'s own configuration layer, which \emph{raises} when an exclusion rule matches nothing rather than silently no-op'ing. A harness is long-running and autonomous; a wrong value that surfaces ten steps downstream costs far more to diagnose than a hard stop at the seam. \emph{Rule.} At every boundary, an unrecognised shape is an error, not a default.

Three observations follow. First, the seams cut across philosophy: fault lines of the same kinds appear in both a batteries-included Python framework and a minimalist TypeScript harness (though not every seam in both, since S1 is Python-specific), suggesting they track the problem rather than either design. Second, the damage concentrates where the seam is irreversible: three of the four specimens end in unrecoverable loss (Table~\ref{tab:seams}, last column), a recursive delete slipping a deny-rule, and conversation history overwritten by compaction, while the one recoverable specimen (S1) merely crashes. Third, the lesson generalises: the expensive seams are exactly the ones an autonomous system cannot walk back, so every automated destructive action (compaction, eviction, deletion) should be recorded and, where possible, reversible. That requirement is precisely what none of the three harnesses provides, and it carries us to the dimension of the next section.

\section{The Boundary: A Dimension Without Convergence (RQ3)}\label{sec:gap}
\begin{figure}[t]
\centering
\includegraphics[width=0.82\linewidth]{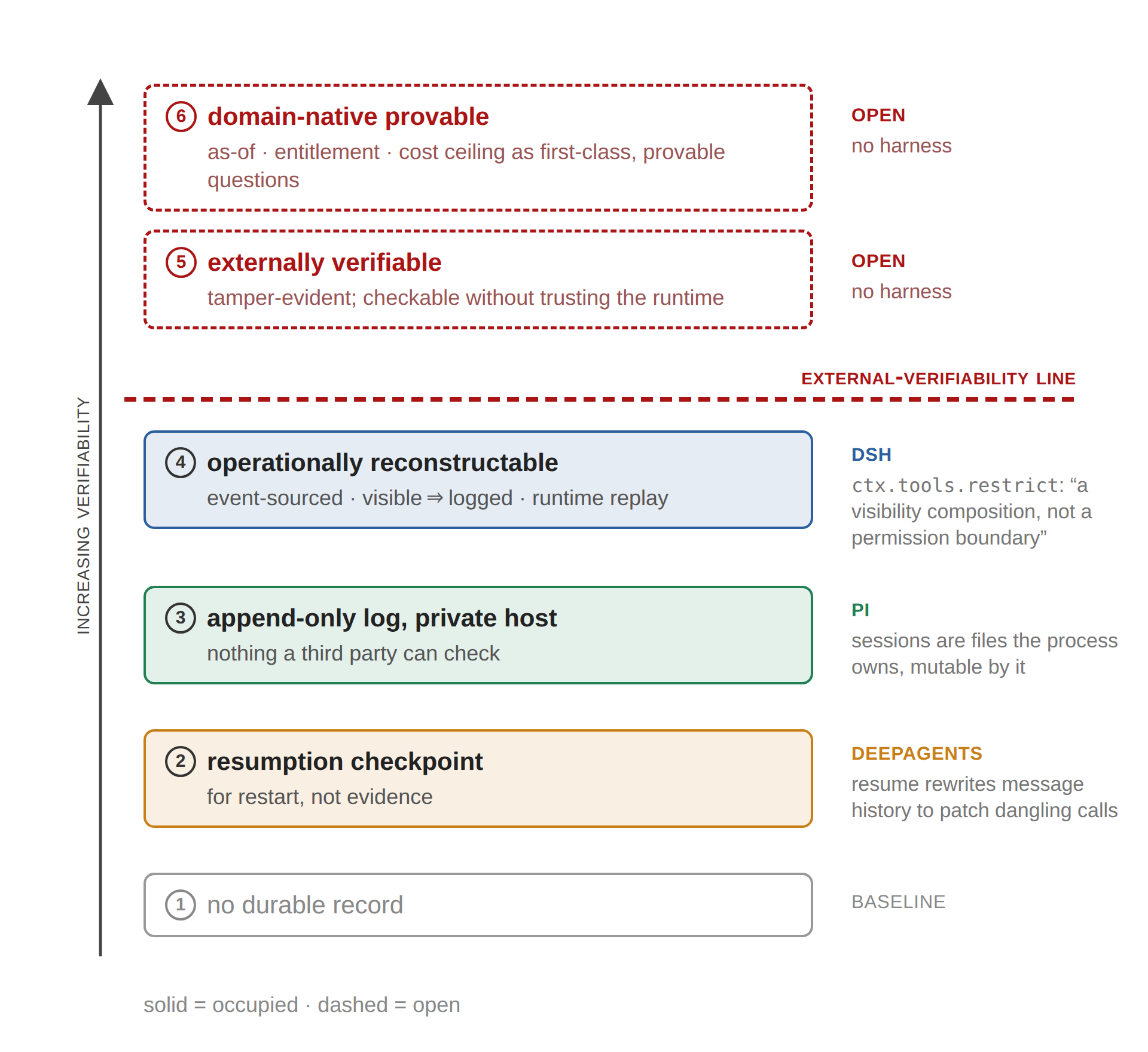}
\caption{\textbf{The verifiability ladder.} \dsh{} climbs highest but stops below the external-verifiability line; the two rungs above it are unreached. Solid\,=\,occupied; dashed\,=\,open. The figure charts the gap only; it does not propose how to fill it.}
\label{fig:ladder}
\end{figure}

This section answers RQ3. Across the dimensions on which the three converge, one load-bearing dimension shows no convergence and, more tellingly, no presence: a record an outside party can verify \emph{without trusting the runtime that produced it}. As shown in Figure~\ref{fig:ladder}, the possibilities form a ladder of increasing verifiability, and every harness stops below the external-verifiability line.

The ladder ascends from a baseline with no durable record. \da{} occupies the rung above it: its checkpoints exist for resumption, not evidence, and its resume path rewrites message history to patch dangling calls. \pipi{} climbs one higher: its sessions are append-only on the newer path, but they remain files the process owns, mutable and unverifiable by a third party. \dsh{} climbs highest and is the instructive exception, demonstrating that the gap is substantive: durable approval records, the visible-implies-logged invariant, recovery that closes rather than truncates. Yet even \dsh{} stops below the line, aiming at \emph{operational} reconstructability (can the runtime rebuild what happened) rather than \emph{external} verifiability, and its own documentation is candid at the boundary: \code{ctx.tools.restrict()} is described as ``a visibility composition, not a permission boundary.'' The two rungs above the line are empty in two distinct ways: no harness produces a tamper-evident record an outsider can check without trusting the runtime (rung~5), and none treats ``what was this data as of,'' ``may this result leave the building,'' or ``prove the cost ceiling held'' as first-class, provable questions (rung~6). The distinction matters because the second rung is the narrower, domain-shaped gap that remains even once generic verifiability exists.

We read this absence as a \emph{predictive gap}, not an oversight. Nor is it absent for lack of proposals: protocol designs such as Autogenesis~\cite{autogenesis} make versioned lineage and auditable rollback their centrepiece, so auditability is actively prescribed top-down in the research literature while absent bottom-up in every harness we read: a selection-pressure asymmetry, not a failure of imagination. The three converged on everything a developer at a terminal needs; verifiability is precisely what that audience does not need, and so precisely where convergence stops. That makes it the natural next axis of competition for harnesses deployed where provenance is not optional. Consistent with our stated discipline, we characterise the gap and do not, here, advance a way to fill it.

\section{Discussion and Implications}
Three implications follow. First, \textbf{for harness builders}, the five elements are a checklist paid for three times under one selection pressure: if you are hand-rolling a loop, mutating session history, coating model quirks in conditionals, front-loading context, or fusing your extensions into a monolith, you are re-deriving a lesson already paid for. We scope that advice to the audience all three serve, a developer at a terminal in an attended, long-horizon session: a scaffold that runs unattended to a benchmark score pays for durability and disclosure it may never draw on, and we would not press the same checklist on it. Within that audience, adopt the settled shapes; do not defend your own version, advice the projects themselves follow, one reusing another's provider catalogue outright. The pull of the form is visible even outside our sample: the reference system accompanying the Autogenesis protocol~\cite{autogenesis}, a research prototype built to demonstrate self-evolution rather than to serve developers, likewise pairs a planner that only plans with sub-agents behind explicit seams, generates capability contracts progressively to spare the prompt, coordinates through a structured \code{plan.md} artefact, and normalises provider quirks in a model manager, a further sighting of the middle form, subject to the same diffusion caveat as any other (Section~\ref{sec:mech}).

Second, \textbf{for the ``less harness'' thesis}, the trajectories qualify it. \da{} is indeed subtracting, which fits ``stronger models need less scaffolding.'' But \pipi{} is simultaneously accreting durable infrastructure (session records, recovery, quirk catalogues) that has nothing to do with coaching the model and everything to do with surviving long autonomous runs. The lesson is not ``thinner everywhere'' but ``thinner in coaching, thicker in durability,'' and how thin is set by audience.

Third, \textbf{for evaluation}, if the harness is the binding constraint~\cite{disclose} and harnesses are converging in form, then the residual performance differences increasingly live in the un-converged dimensions (recovery semantics, quirk coverage, and the verifiability others have not built) rather than in the loop everyone now shares. For future harness work, the implication is to adopt the settled shapes, differentiate on the unsettled dimensions, and treat the verifiable record as the likely next requirement where deployment domains demand provenance.

\section{Threats to Validity}\label{sec:threats}
\textbf{External validity.} $N=3$; all are coding-agent harnesses read at one point in time (August 2026), so generalisation to other agent shapes and to future revisions is limited. We mitigate by maximal-spread selection and by pinning revisions so claims are re-checkable.

\textbf{Internal validity.} The three are not fully independent (\dsh{} depends on \pipi{}), so part of the convergence is diffusion, not rediscovery; Section~\ref{sec:mech} handles this by decomposing the mechanism rather than assuming independence. Case selection favours well-known, source-available projects, a survivorship bias we state plainly.

\textbf{Construct validity.} Two constructs carry different risk. The \emph{dimensions} (Table~\ref{tab:dims} rows) were fixed a priori (Section~\ref{sec:design}) and aligned with a prior taxonomy~\cite{scaffold}, limiting fitting. The \emph{five elements} (the central construct) were instead induced during reading of the two grounding cases and checked against one more, and we mark this as the primary construct risk; the inclusion rule was line-level evidence in both grounding cases. We also disclose that the verifiability dimension aligns with the first author's own research interest, stated here so the reader can weight it. The upstream-merged fix confirms one point claim, not the framework; we scope it accordingly (Section~\ref{sec:design}).

\textbf{Reliability.} A single analyst read the source with AI assistance, which introduces both single-coder and instrument risk; \dsh{} was read less deeply than the other two, and we mark \dsh{} claims as documentation-level where we did not reach line evidence. Against this, every file:line is re-checkable at the pinned revisions and the reproduction artefacts are executable. A second-coder re-scoring of a sample of table cells, and inter-run agreement for the AI-assisted pass, are the natural next mitigations and are deferred to the scaled study. 
That study is underway as a companion paper, which re-expresses the dimensions of Table~\ref{tab:dims} as blind detectors bound by a mandatory file:line evidence contract, applies them in independent replicated runs across a stated population of harnesses, and so reports inter-run agreement and detector-versus-human agreement as measurements rather than as promises. The present paper's labels are the calibration set that study is checked against; we therefore state them here without borrowing its results.

\section{Conclusion}
Three harnesses began from opposing philosophies and evolved in opposite directions, yet moved toward one middle form: a commoditised loop, an append-only replayable record, quirks as data, progressive disclosure, and explicit seams. The convergence is real but not independent: we saw parallel discovery, diffusion, and one case of literal reuse. When separate teams under the same pressure keep landing on the same shapes, those shapes are settled, and the engineering move is to adopt them rather than defend one's own. The same current that reveals the settled parts throws the unsettled one into relief: external verifiability, a record an outsider can check without trusting the runtime, is a dimension on which all three are not merely divergent but absent. That is where the next harness will differ from the three we read.

\section*{Data Availability}
Reproduction artefacts (the two \dsh{} plugins and the two sandbox defect reproductions) and the full per-dimension evidence table with file:line references are available in the supplementary material, to be archived with a DOI on release; the defect reproductions are released under coordinated disclosure. All source claims are verifiable at the pinned revisions and repositories of Section~\ref{sec:design}.

\section*{Acknowledgement of AI Assistance}
The source sweep underlying this study used parallel AI agents to locate and excerpt code; every claim was subsequently verified by hand against the source by the author, who is solely responsible for all errors.


\end{document}